\documentclass[conference]{IEEEtran}
\IEEEoverridecommandlockouts
\usepackage{lineno} 
\usepackage{cite}
\usepackage{amsmath,amssymb,amsfonts}
\usepackage{algorithmic}
\usepackage{graphicx}
\usepackage{textcomp}
\usepackage{graphics}
\usepackage{listings}
\usepackage{amsmath}
\usepackage{textcomp}
\usepackage{pgfplots}
\usepackage{pgf-pie}
\pgfplotsset{width=7cm,compat=1.5}
\usepackage{multirow}
\usepackage{tikz}
\usepackage{xcolor} 
\usepackage{framed} 
\usepackage{enumitem}
\definecolor{rose}{RGB}{236,64,122}
\definecolor{gold}{RGB}{179,134,0}
\definecolor{shadecolor}{rgb}{0.9,0.9,0.9} 
\usepackage{xspace}
\usepackage{graphicx}
\usepackage{ifthen}
\usepackage[normalem]{ulem} 
\usepackage{xcolor,colortbl}
\usepackage{hyperref}

\newboolean{showedits}
\setboolean{showedits}{true} 
\ifthenelse{\boolean{showedits}}
{
	\newcommand{\del}[1]{\textcolor{red}{\sout{#1}}} 
	\newcommand{\nbe}[3]{
		{\colorbox{#3}{\bfseries\sffamily\scriptsize\textcolor{white}{#1}}}
		{\textcolor{#3}{\sf\small$\blacktriangleright$\textit{#2}$\blacktriangleleft$}}}
}{
	\newcommand{\del}[1]{} 
	
	\newcommand{\nbe}[3]{}
}

\newboolean{showcomments}
\setboolean{showcomments}{true} 
\newcommand{\id}[1]{$-$Id: scgPaper.tex 32478 2010-04-29 09:11:32Z oscar $-$}

\ifthenelse{\boolean{showcomments}}
 {
 	\newcommand{\nbc}[3]{
 		{\colorbox{#3}{\bfseries\sffamily\scriptsize\textcolor{white}{#1}}}
		{\textcolor{#3}{\sf\small$\blacktriangleright$\textit{#2}$\blacktriangleleft$}}}
	
 }{
 	\newcommand{\nbc}[3]{}
 	
 }

\usepackage[most]{tcolorbox}
\ifthenelse{\boolean{showedits}}
{
  \newtcolorbox{inserted}{%
       title=Inserted text:,
       colframe=blue,colback=blue!5!white,
       breakable,
       leftrule=0mm, 
       bottomrule=0mm,
       rightrule=0mm,
       toprule=0mm,
       arc=0mm, outer arc=0mm,
       oversize
  }
  \newtcolorbox{deleted}{%
       title=Deleted text:,
       colframe=red,colback=red!5!white,
       breakable,
       leftrule=0mm, 
       bottomrule=0mm,
       rightrule=0mm,
       toprule=0mm,
       arc=0mm, outer arc=0mm,
       oversize
  }
  \newtcolorbox{refactored}{%
       title=Rewritten text:,
       colframe=blue,colback=red!5!white,
       breakable,
       leftrule=0mm, 
       bottomrule=0mm,
       rightrule=0mm,
       toprule=0mm,
       arc=0mm, outer arc=0mm,
       oversize
  }
}{

}
\newboolean{isblinded}
\setboolean{isblinded}{true}
\ifthenelse{\boolean{isblinded}}
{\newcommand\blind[1]{BLINDED\xspace}}
{\newcommand\blind[1]{#1\xspace}}

\newcommand{\commented}[1]{}

\definecolor{source}{gray}{0.9}

\lstdefinelanguage{Java}{
  tabsize=4
}[keywords,comments,strings]

\definecolor{source}{gray}{0.95}
\definecolor{highlight}{gray}{0.9}
\definecolor{bblue}{HTML}{4F81BD}
\definecolor{rred}{HTML}{C0504D}
\definecolor{ggreen}{HTML}{9BBB59}
\definecolor{ppurple}{HTML}{9F4C7C}
\usetikzlibrary{patterns}
\usepackage{tabularx}

\usepackage{multirow}
\usepackage{booktabs} 
\usepackage{siunitx}
\usepackage{caption}  
\usepackage{fontawesome5}
\usepackage{colortbl}
\usetikzlibrary{matrix}
\usepackage{colortbl}
\usepackage{pgfplotstable}

\pgfplotsset{compat=1.18}

\usepackage[most]{tcolorbox}
\newcommand{\boxit}[2][yellow!20]{%
\begin{tcolorbox}[colback=#1,  boxrule=0.5pt, arc=1mm, breakable]
#2
\end{tcolorbox}
}
\usepackage[T1]{fontenc}
\usepackage{pgfplotstable}

\lstnewenvironment{codesnippet}{%
	\lstset{%
    float,
		frame=single,
		framerule=0pt,
		mathescape=false
	}
}{}
\def\BibTeX{{\rm B\kern-.05em{\sc i\kern-.025em b}\kern-.08em
    T\kern-.1667em\lower.7ex\hbox{E}\kern-.125emX}}

    \lstdefinestyle{customStyle}{
  basicstyle=\ttfamily\scriptsize,  
  breaklines=true,                  
  frame=none,                       
  backgroundcolor=\color{white},    
  showstringspaces=false,           
  numbers=none                      
  }

\begin{document}
\title{
Persistent Human Feedback, LLMs, and Static Analyzers for Secure Code Generation and Vulnerability Detection
}

\author{
  \IEEEauthorblockN{Ehsan Firouzi}
  \IEEEauthorblockA{Technische Universität Clausthal\\
                    Germany
                   }
  \and
  \IEEEauthorblockN{Mohammad Ghafari}
  \IEEEauthorblockA{Technische Universität Clausthal\\
                   Germany
                   }
}

\maketitle

\begin{abstract}

Existing literature heavily relies on static analysis tools to evaluate LLMs for secure code generation and vulnerability detection. 
We reviewed 1,080 LLM-generated code samples, built a human-validated ground-truth, and compared the outputs of two widely used static security tools, CodeQL and Semgrep, against this corpus. 
While 61\% of the samples were genuinely secure, Semgrep and CodeQL classified 60\% and 80\% as secure, respectively. 
Despite the apparent agreement in aggregate statistics, per-sample analysis reveals substantial discrepancies: only 65\% of Semgrep’s and 61\% of CodeQL’s reports correctly matched the ground truth. 
These results question the reliability of static analysis tools as sole evaluators of code security and underscore the need for expert feedback. Building on this insight, we propose a conceptual framework that persistently stores human feedback in a dynamic retrieval-augmented generation pipeline, enabling LLMs to reuse past feedback for secure code generation and vulnerability detection.

\end{abstract}


\section{Introduction}
\label{sec:Intro}

Large language models ( LLMs) have turned into an important part of everyday software development. The 2025 Stack Overflow Developer Survey reports that 82\% of developers used OpenAI’s GPT models in their work over the last year~\cite{stackoverflow_ai_2025}. This widespread adoption highlights both the promise and the risks of relying on LLMs for code generation.

Despite their practical benefits, LLMs can generate insecure code and configurations even when explicitly instructed to produce secure solutions~\cite{Firouzi2026SecureIaC}, undermining trust in their results. Static analysis tools can effectively surface potential vulnerabilities; however, research shows that they may exhibit high false-positive rates~\cite{Crypto-misuses} and can be difficult to act upon without additional context. Recent studies also highlight LLMs’ ability to identify certain classes of vulnerabilities\cite{FIROUZIGenAI,FirouziChatGPTPotential,FIROUZItime}.

In this paper, we first survey publications from top-tier software engineering venues in 2025 to characterize current research on the use of LLMs for secure code generation and vulnerability detection. We examine how the literature evaluates generated code, including the benchmarks used, target programming languages, and evaluation practices. Our analysis shows that existing benchmarks predominantly focus on Python for code generation and C/C++ for vulnerability detection, while also exhibiting divergent CWE distributions. In addition, we observe a strong reliance on static analysis tools rather than human inspection.

To assess the reliability of automated evaluation tools, we manually reviewed 1,080 GPT-4o–generated code samples produced using the RCI (Recursive Critiques Improvement) prompting technique, which prior work has shown to yield more secure outputs~\cite{Bruni2025}. We constructed a ground-truth dataset of vulnerable and secure code and compared it against the reports generated by two widely used static security analysis tools, CodeQL and Semgrep. Our manual inspection found that 61\% of the samples were genuinely secure. In comparison, Semgrep and CodeQL classified 60\% and 80\% of the samples as secure, respectively. Although Semgrep’s overall classification rate appears closely aligned with human judgment, a per-sample analysis reveals substantial disagreement: only 65\% of Semgrep’s reports and 61\% of CodeQL’s reports matched the ground-truth labels.

These findings raise concerns about the reliability of static analysis tools when used as the sole means of evaluating code security and underscore the continued importance of expert feedback. 
Building on this observation, we find that expert feedback is important not only for individual instances but also may generalize to similar cases. Accordingly, we propose a conceptual framework that persistently stores human feedback within a dynamic retrieval-augmented generation pipeline, enabling LLMs to reuse prior feedback for secure code generation and vulnerability detection.

The rest of this paper is structured as follows.
In Section \ref{sec:litreture}, We review the recent literature on the use of LLMs for code security. In Section \ref{sec:study2}, we describe our study and present the results. 
In Section \ref{sec:framework}, we introduce our conceptual framework to persist human feedback.
We discuss threats to validity of our study in Section \ref{sec:threats} and conclude the paper in Section \ref{sec:Conclusion}.

\section{Literature Review}
\label{sec:litreture}

We surveyed papers published in 2025 in top-tier (A/A*) software engineering venues and reviewed those that investigate the use of LLMs for secure code generation and vulnerability detection. In particular, we collected information on the evaluation approaches, the benchmarks employed, and the tools used.


\begin{table*}[ht]
\centering
\caption{Overview of studies on secure code generation
}
\scalebox{0.73}{
\begin{tabular}{|l|l|l|l|l|l|l|l|}
\hline

\textbf{Study} &\textbf{Venue} &\textbf{Benchmarks} & \textbf{Languages} &\textbf{Model Families} & \textbf{Evaluation Tools} & \textbf{Manual Inspection}  & \textbf{RAG}\\ \hline

\begin{tabular}[c]{@{}l@{}}Towards Secure Code Generation \\ with LLMs: A Study on Common \\ Weakness Enumeration\cite{zhao2025towards}\end{tabular} & TSE & \multicolumn{1}{l|}{LLMSecEval} & \multicolumn{1}{c|}{\begin{tabular}[c]{@{}c@{}}Python \\ C\end{tabular}} & \multicolumn{1}{l|}{\begin{tabular}[c]{@{}l@{}}OpenAI GPT \\ Anthropic Claude \\ Meta Llama/CodeLlama \\ Qwen2.5 Coder \\ DeepSeek Coder\end{tabular}} & \multicolumn{1}{l|}{No} & Yes  & Yes \\ \hline

\begin{tabular}[c]{@{}l@{}}Security Weaknesses of Copilot-Generated \\ Code in GitHub Projects: An Empirical Study\cite{fu2025security}\end{tabular} & TOSEM & \multicolumn{1}{l|}{\begin{tabular}[c]{@{}l@{}}AI-generated \\ code from \\ Public GitHub\end{tabular}} & \multicolumn{1}{c|}{\begin{tabular}[c]{@{}c@{}}Python \\ JavaScript\end{tabular}} & \multicolumn{1}{l|}{\begin{tabular}[c]{@{}l@{}}OpenAI GPT \\ (via GitHub Copilot Chat)\end{tabular}} & \multicolumn{1}{l|}{\begin{tabular}[c]{@{}l@{}}CodeQL \\ Bandit \\ ESLint\end{tabular}} & Yes & No \\ \hline
\begin{tabular}[c]{@{}l@{}}Prompting Techniques for Secure Code \\ Generation: A Systematic Investigation\cite{tony2025prompting}\end{tabular} & TOSEM & \multicolumn{1}{l|}{LLMSecEval} & \multicolumn{1}{c|}{Python} & \multicolumn{1}{l|}{OpenAI GPT} & \multicolumn{1}{l|}{\begin{tabular}[c]{@{}l@{}}Bandit \\ CodeQL\end{tabular}} & No & No \\ \hline

{\color[HTML]{000000}\begin{tabular}[c]{@{}l@{}}Discrete Prompt Optimization Using Genetic \\ Algorithm for Secure Python Code Generation\cite{tony2025discrete}\end{tabular}} & {\color[HTML]{000000}JSS} & \multicolumn{1}{l|}{{\color[HTML]{000000}\begin{tabular}[c]{@{}l@{}}LLMSecEval \\ SecurityEval\end{tabular}}} & \multicolumn{1}{c|}{{\color[HTML]{000000}Python}} & \multicolumn{1}{l|}{{\color[HTML]{000000}\begin{tabular}[c]{@{}l@{}}OpenAI GPT \\ Google Gemini \\ Meta CodeLlama \\ DeepSeek Coder\end{tabular}}} & \multicolumn{1}{l|}{{\color[HTML]{000000}\begin{tabular}[c]{@{}l@{}}Bandit \\ CodeQL\end{tabular}}} & {\color[HTML]{000000}Yes} & No \\ \hline

\begin{tabular}[c]{@{}l@{}}Retrieve, Refine, or Both? Using Task-Specific \\ Guidelines for Secure Python Code Generation\cite{tony2025retrieve}\end{tabular} & ICSME & \multicolumn{1}{l|}{LLMSecEval} & \multicolumn{1}{c|}{Python} & \multicolumn{1}{l|}{\begin{tabular}[c]{@{}l@{}}OpenAI GPT \\ Google Gemini \\ DeepSeek Coder\end{tabular}} & \multicolumn{1}{l|}{\begin{tabular}[c]{@{}l@{}}Bandit \\ CodeQL\end{tabular}} & No & Yes  \\ \hline

\begin{tabular}[c]{@{}l@{}}Evaluating Software Development Agents: \\ Patch Patterns, Code Quality, and Issue \\ Complexity in Real-World GitHub Scenarios\cite{chen2025evaluating}\end{tabular} & SANER & \multicolumn{1}{l|}{SWE-Bench} & \multicolumn{1}{c|}{Python} & \multicolumn{1}{l|}{\begin{tabular}[c]{@{}l@{}}Multiple families \\ (various LLM backends)\end{tabular}} & \multicolumn{1}{l|}{SonarQube} & No & No  \\ \hline

\begin{tabular}[c]{@{}l@{}}Is LLM-Generated Code More Maintainable \\ \& Reliable than Human-Written Code?\cite{molison2025llm}\end{tabular} & ESEM & \multicolumn{1}{l|}{SWE-Bench} & \multicolumn{1}{c|}{Python} & \multicolumn{1}{l|}{OpenAI GPT (4o)} & \multicolumn{1}{l|}{SonarQube} & \multicolumn{1}{l|}{Yes  } & No \\ \hline
\begin{tabular}[c]{@{}l@{}}Do Prompt Patterns Affect Code Quality? \\ A First Empirical Assessment of ChatGPT-Generated Code\cite{della2025prompt}\end{tabular} & EASE & \multicolumn{1}{l|}{DevGPT} & \multicolumn{1}{c|}{Various} & \multicolumn{1}{l|}{OpenAI GPT (4o)} & \multicolumn{1}{l|}{SonarQube} & No & No  \\ \hline
\begin{tabular}[c]{@{}l@{}}How Well Do Large Language Models Serve \\ as End-to-End Secure Code Agents for Python?~\cite{gong2024well}\end{tabular} & EASE & \multicolumn{1}{l|}{SecurityEval} & \multicolumn{1}{c|}{Python} & \multicolumn{1}{l|}{\begin{tabular}[c]{@{}l@{}}OpenAI GPT \\ Meta CodeLlama \\ CodeGeeX2\end{tabular}} & \multicolumn{1}{l|}{\begin{tabular}[c]{@{}l@{}}Bandit \\ CodeQL\end{tabular}} & Yes & No  \\ \hline
\end{tabular}
}
\label{tbl:llm_secure_code_gen_overview}
\end{table*}

\begin{table*}[ht]
\centering

\caption{Overview of studies on vulnerability detection} 
{\raggedright\scriptsize
{Only tools marked with \faBug~ are used as security checkers; the others are used to enhanced inputs (e.g., AST/CFG construction for pre-processing).}}\\[2mm]
\scalebox{0.8}{
\begin{tabular}{|m{6.2cm}|m{0.9cm}|m{3.3cm}|m{1.9cm}|m{3.7cm}|p{2cm}|m{1.3cm}|m{1cm}|}
\hline
\textbf{Study} & \textbf{Venue} & \textbf{Benchmarks/Datasets} & \textbf{Language} & \textbf{Model Families} & \textbf{Analysis Tools} & \textbf{Manual Inspection} & \textbf{RAG} \\ \hline
Vulnerability Detection with Code Language Models: How Far Are We? \cite{ICSEding2024vulnerability} & ICSE & PRIMEVUL & C/C++ & CodeBERT; CodeT5; UniXcoder; StarCoder; CodeGen; OpenAI GPT &\-- & Yes & No \\ \hline
GVI: Guided Vulnerability Imagination for Boosting Deep Vulnerability Detectors\cite{ICSEyong2025gvi} & ICSE & Devign, ReVeal, Big-Vul & C/C++ & OpenAI GPT & \--  & No & No \\ \hline
CGP-Tuning: Structure-Aware Soft Prompt Tuning for Code Vulnerability Detection\cite{TSE_CGP-Tuning} & TSE & DiverseVul & C/C++ & Meta Llama/CodeLlama; Google Gemma/CodeGemma & Joern & No & No \\ \hline
Improving Co-Decoding Based Security Hardening of Code LLMs leveraging knowledge distillation \cite{li2025improving} & TSE & Evol-CodeAlpaca, Security Training Dataset & Py, C++ & CodeGen; StarCoder & \--  & No & No \\ \hline
SecureFalcon: Are We There Yet in Automated Software Vulnerability Detection with llms?\cite{ferrag2025securefalcon} & TSE & FormAI, FalconVulnDB & C/C++ & SecureFalcon (custom) & \--  & Yes & No \\ \hline
{Towards Explainable Vulnerability Detection With Large Language Models} \cite{mao2025towards} & TSE & SeVC, DiverseVul & C/C++ & Meta Llama/CodeLlama; DeepSeek &\--  & Yes & No \\ \hline
{VulScribeR: Exploring RAG-based vulnerability augmentation with LLMs}\cite{Tdaneshvar2025vulscriber} & TOSEM & Devign, BigVul, Reveal, PrimeVul & C/C++ & OpenAI GPT; Qwen/CodeQwen; CodeBERT & Joern  & Yes & Yes \\ \hline
{On the Evaluation of Large Language Models in Multilingual Vulnerability Repair}\cite{Tchen2025security} & TOSEM & REEF & C, C++, C\#, Go, Java, JS, Py & OpenAI GPT; DeepSeek Coder; Meta CodeLlama; Meta Llama 3 & Tree-sitter & Yes & No \\ \hline
{Learning-based models for vulnerability detection: an extensive study} \cite{ni2026learning} & EMSE & MegaVul & C/C++ & OpenAI GPT; LineVul & Joern & No & No \\ \hline
{VulTrLM: LLM-assisted vulnerability detection via AST decomposition and comment enhancemen}\cite{zhang2026vultrlm} & EMSE & FFmpeg+QEMU (Devign), Reveal, SVulD & C/C++ & Qwen/Qwen Coder; DeepSeek; UniXcoder & Tree-sitter & No & No \\ \hline
{A zero-shot framework for cross-project vulnerability detection in source code}\cite{haque2026zero} & EMSE & Devign, REVEAL & C/C++ & CodeBERT; UniXcoder & \-- & No & No \\ \hline
{Dlap: A deep learning augmented large language model prompting framework for software vulnerability detection}\cite{JSSyang2025dlap} & JSS & Chrome, Linux, Android, QEMU & C/C++ & OpenAI GPT; Vicuna (LLaMA-based) & Flawfinder ~\faBug,  Cppcheck ~\faBug, Joern  & No & No \\ \hline
{Enhancing vulnerability repair through the extraction and matching of repair patterns}\cite{JSScao2025enhancing} & JSS & CVEFixes + Big-Vul & C/C++ & OpenAI GPT & \-- & No & No \\ \hline
{Improving distributed learning-based vulnerability detection via multi-modal prompt tuning}\cite{JSSren2025improving} & JSS & Devign, ReVeal, Big-Vul & C/C++ & GraphCodeBERT & \-- & No & No \\ \hline
{Transfer learning for software vulnerability prediction using Transformer models}\cite{JSSkalouptsoglou2025transfer} & JSS & Big-Vul, FFmpeg+QEMU & C/C++ & CodeBERT; CodeGPT & \-- & No & No \\ \hline
{Mystique: Automated Vulnerability Patch Porting with Semantic and Syntactic-Enhanced LLM} \cite{wu2025mystique} & FSE & PatchDB & C/C++ & StarCoder & Joern, Tree-sitter, Clang-tidy & Yes & No \\ \hline
{Large Language Models for In-File Vulnerability Localization Can Be “Lost in the End”}\cite{FSELostintheEnd} & FSE & MITRE CVE catalog & C/C++ & Mistral Mixtral; Meta Llama 3; OpenAI GPT &\-- & Yes & No \\ \hline
{One-for-All Does NotWork! Enhancing Vulnerability Detection by Mixture-of-Experts}\cite{FSEMoE} & FSE & BigVul & C/C++ & CodeBERT; UniXcoder &\-- & No & No \\ \hline
{Efficient and robust security-patch localization for disclosed oss vulnerabilities with fine-tuned llms in an industrial setting}\cite{FSEran2025efficient} & FSE & CVEfixes , Huawei Cloud internal curation (303 CVEs) & Java, C++ & Qwen; Meta Llama 3.1 & \-- & Yes & No \\ \hline
{Vul-R2: A Reasoning LLM for Automated Vulnerability Repair} \cite{wen2025vul} & ASE & PrimeVul, SVEN & C/C++ & Qwen; OpenAI o3 & \-- & Yes & No \\ \hline
{SIExVulTS: Sensitive Information Exposure Vulnerability Detection System using Transformer Models and Static Analysis}\cite{ESEMkatz2025siexvults} & ESEM & Synthetic CWE-200 dataset, CVE dataset (real-world), SIEx-Flow & Java & OpenAI GPT; Sentence-BERT; CodeBERT; CodeT5; GraphCodeBERT &  CodeQL ~\faBug & Yes & No \\ \hline
{LLM-SZZ: Novel Vulnerability-Inducing Commit Identification Driven by Large Language Model and CVE Description}\cite{fanllm} & ICSME & SZZ Benchmarks & Java & OpenAI GPT; DeepSeek; Mistral Mixtral & \-- & Yes & No \\ \hline
{Enhanced Vulnerability Localization: Harmonizing Task-Specific Tuning and General LLM Prompting}\cite{tianenhanced} & ICSME & SVEN, BigVul, Devign & C/C++ & Meta CodeLlama; Anthropic Claude; DeepSeek R1 & \-- & No & No \\ \hline
{Repairing vulnerabilities without invisible hands. A differentiated replication study on LLMs}\cite{ICSMEcamporese2025repairing} & ICSME & Vul4J, VJBench & Java & OpenAI GPT; DeepSeek R1; Qwen & \-- & Yes & No \\ \hline

\end{tabular}}
\\
\label{tbl:llm_secure_VulDetect_overview}
\end{table*}

\subsection{Observations}

Table~\ref{tbl:llm_secure_code_gen_overview} summarizes the reviewed studies on secure code generation, and Table~\ref{tbl:llm_secure_VulDetect_overview} presents those focused on vulnerability detection. The \emph{study} column in these tables lists the paper titles, while \emph{benchmarks/datasets} specifies the evaluation benchmarks used in the study and \emph{language} reports the target programming languages. We further record the \emph{model families} evaluated or used, the \emph{evaluation/analysis tools} (i.e., tools used either to enhance inputs or to support evaluation), whether \emph{manual inspection} of model outputs was conducted by human experts, and whether the method incorporates \emph{retrieval-augmented generation (RAG)} to ground the model on external security-relevant information.


\begin{table}[ht]
\centering
\caption{
List of analysis tools}
\raggedright\scriptsize
\faBug: Used as a security checker\\
\faCogs: Used to enhance inputs (e.g., AST/CFG construction for pre-processing)\\[3mm]
\label{tbl:static_tools_overview}
\scalebox{0.85}{
\begin{tabular}{|l|c|m{1.5cm}|m{3.5cm}|m{0.7cm}|}
\hline
\textbf{Tool Name}& \textbf{Purpose} & \textbf{Language Support} & \textbf{Studies} & \textbf{Count} \\ \hline
\textbf{CodeQL} & \faBug &Multi (Py, Java, C++) & ~\cite{fu2025security}, ~\cite{tony2025prompting}, ~\cite{tony2025discrete}, ~\cite{tony2025retrieve}, ~\cite{gong2024well} & 5 \\ \hline
\textbf{Bandit} & \faBug & Python & ~\cite{fu2025security}, ~\cite{tony2025prompting}, ~\cite{tony2025retrieve}, ~\cite{gong2024well} & 4 \\ \hline
\textbf{SonarQube}& \faBug & Multi & ~\cite{chen2025evaluating}, ~\cite{molison2025llm}, ~\cite{della2025prompt} & 3 \\ \hline
\textbf{ESLint}& \faBug & JavaScript & ~\cite{fu2025security} & 1 \\ \hline
\textbf{Joern} &\faCogs & C/C++ & 
~\cite{TSE_CGP-Tuning}, 
~\cite{Tdaneshvar2025vulscriber}, 
~\cite{ni2026learning}, 
~\cite{JSSyang2025dlap}, 
~\cite{wu2025mystique} & 5 \\ \hline
\textbf{Tree-sitter}&\faCogs & Multi & 
~\cite{Tchen2025security}, 
~\cite{zhang2026vultrlm}, 
~\cite{wu2025mystique} & 3 \\ \hline
\textbf{Flawfinder}&\faCogs & C/C++ & 
~\cite{JSSyang2025dlap} & 1 \\ \hline
\textbf{Cppcheck}&\faCogs & C/C++ & 
~\cite{JSSyang2025dlap} & 1 \\ \hline
\textbf{CodeQL}&\faCogs & Multi (Py, Java, C++) & 
~\cite{ESEMkatz2025siexvults} & 1 \\ \hline
\textbf{Bandit}&\faCogs & Python & ~\cite{tony2025discrete} & 1 \\ \hline

\end{tabular}
}
\\
\end{table}

\emph{Benchmarks and datasets}. Table \ref{tbl:benchmark_taxonomy} outlines the benchmarks used across these studies. Studies on vulnerability detection rely more on datasets like Devign, ReVeal, Big-Vul, PrimeVul\cite{ICSEding2024vulnerability}, and MegaVul\cite{ni2026learning} (C/C++).
~In secure code generation, LLMSecEval~\cite{10174231} and SecurityEval \cite{siddiq2022securityeval} (Python) are commonly used; however, when it comes to evaluating code quality more than security, researchers often use SWE-bench.
We also observed divergent CWE distributions on benchmarks, which can lead to misleading conclusions in empirical studies.

\begin{table}[ht]
\centering
\caption{
Benchmarks and Datasets
}

\raggedright\scriptsize
\faShieldVirus: Vulnerability Detection and Repair Benchmarks\\
\faCode: Secure Code Generation Benchmarks \\[3mm]
\label{tbl:benchmark_taxonomy}
\scalebox{0.80}{
\begin{tabular}{|l|c|m{1.5cm}|m{3.6cm}|m{0.7cm}|}
\hline
\textbf{Benchmark / Dataset} & \textbf{Purpose}  &\textbf{Language Support} & \textbf{Studies} & \textbf{Count} \\ \hline
\textbf{Big-Vul} & \faShieldVirus & C/C++ & \cite{ICSEyong2025gvi}, ~\cite{Tdaneshvar2025vulscriber}, ~\cite{JSScao2025enhancing}, ~\cite{JSSren2025improving}, ~\cite{JSSkalouptsoglou2025transfer}, ~\cite{FSEMoE}, ~\cite{tianenhanced} & 7 \\ \hline
\textbf{Devign} & \faShieldVirus & C/C++ & ~\cite{ICSEyong2025gvi}, ~\cite{Tdaneshvar2025vulscriber}, ~\cite{zhang2026vultrlm}, ~\cite{haque2026zero}, ~\cite{JSSren2025improving}, ~\cite{tianenhanced} & 6 \\ \hline
\textbf{ReVeal} & \faShieldVirus & C/C++ & ~\cite{ICSEyong2025gvi}, ~\cite{Tdaneshvar2025vulscriber}, ~\cite{zhang2026vultrlm}, ~\cite{haque2026zero}, ~\cite{JSSren2025improving} & 5 \\ \hline
\textbf{PrimeVul} & \faShieldVirus & C/C++ & ~\cite{ICSEding2024vulnerability}, ~\cite{Tdaneshvar2025vulscriber}, ~\cite{wen2025vul} & 3 \\ \hline
\textbf{DiverseVul} & \faShieldVirus & C/C++ & ~\cite{TSE_CGP-Tuning}, ~\cite{mao2025towards} & 2 \\ \hline
\textbf{SVEN}& \faShieldVirus  & C/C++ & ~\cite{wen2025vul}, ~\cite{tianenhanced} & 2 \\ \hline
\textbf{CVEFixes} & \faShieldVirus & Multi & ~\cite{JSScao2025enhancing}, ~\cite{FSEran2025efficient} & 2 \\ \hline
\textbf{Vul4J / VJBench} & \faShieldVirus & Java & ~\cite{ICSMEcamporese2025repairing} & 1 \\ \hline
\textbf{PatchDB} & \faShieldVirus & C/C++ & ~\cite{wu2025mystique} & 1 \\ \hline
\textbf{MegaVul} & \faShieldVirus & C/C++ & ~\cite{ni2026learning} & 1 \\ \hline
\textbf{FormAI / FalconVulnDB} & \faShieldVirus & C/C++ & ~\cite{ferrag2025securefalcon} & 1 \\ \hline
\textbf{LLMSecEval} & \faCode & Py, C & ~\cite{zhao2025towards}, ~\cite{tony2025prompting}, ~\cite{tony2025discrete}, ~\cite{tony2025retrieve} & 4 \\ \hline
\textbf{SecurityEval}  & \faCode & Python & ~\cite{tony2025discrete}, ~\cite{gong2024well} & 2 \\ \hline
\textbf{SWE-Bench} & \faCode & Python & ~\cite{chen2025evaluating}, ~\cite{molison2025llm} & 2 \\ \hline
\end{tabular}}
\\
\end{table}

\emph{Programming languages}. Studies targeting vulnerability detection primarily focus on C/C++, where memory safety issues such as buffer overflows and pointer mismanagement are prevalent, with Java remaining a distant second. Conversely, studies on secure code generation predominantly evaluate Python. This discrepancy creates a significant knowledge gap regarding multilingual secure coding capabilities.

\emph{Models}. Table \ref{tbl:model_landscape} categorizes the language models employed. The GPT family (GPT-3.5/4) remains the dominant closed-source baseline across all tasks. Among open-source models, Llama/CodeLlama is the most frequent target for fine-tuning experiments, while CodeBERT retains its status as the standard encoder-only model for embedding-based detection tasks.

\emph{Evaluation tools}. Table \ref{tbl:static_tools_overview} presents the tools utilized in recent studies. Their application differs by domain: in secure code generation, static analysis tools are primarily used as evaluation oracles (e.g., CodeQL) to score the security of the final output, while in vulnerability detection, tools are frequently employed for guidance and preprocessing (e.g., Joern), extracting features such as abstract syntax trees (ASTs) or control flow graphs (CFGs) to enhance model understanding.

\emph{Manual Analysis}.
Existing studies exhibit low human oversight and heavily rely on automated static analysis tools. Although scalable, this approach risks missing vulnerabilities that tools cannot detect and introducing false positives.
Even in studies that incorporate manual validation, authors rarely explain how tool-based and human assessments differ in practice. For example, Fu et al. \cite{fu2025security} note that two reviewers manually inspected filtered insecure code and achieved $\kappa = 0.82 $, but do not quantify discrepancies between automated and manual results. Similarly, Tony et al. \cite{tony2025discrete} only state that Bandit and CodeQL warnings were manually reviewed and false positives discarded, and Gong et al. \cite{gong2024well} simply mark a sample as vulnerable if any of manual review, CodeQL, or Bandit flags it as insecure.

\boxit{
Most existing work relies heavily on automated static analysis with limited human oversight, and even when manual review is included, 
little insight exists into the extent to which human and tool assessments diverge.}


\renewcommand{\arraystretch}{1.25} 
\begin{table}[ht]
\centering
\caption{List of LLMs}
\label{tbl:model_landscape}

\setlength{\tabcolsep}{8pt} 

\scalebox{0.68}{
\begin{tabular}{|l|p{5cm}|m{4.7cm}|c|}
\hline
\multicolumn{2}{|c|}{\textbf{Model}} & \textbf{Studies} & \textbf{Count} \\ \hline

\multirow{5}{*}{\rotatebox[origin=c]{90}{\parbox{1.7cm}{Closed Source}}}%
  & \textbf{GPT-4 Family} (GPT-4, 4o, 4o-mini, Turbo) &
    \cite{zhao2025towards}, \cite{fu2025security}, \cite{tony2025prompting},
    \cite{tony2025discrete}, \cite{tony2025retrieve}, \cite{molison2025llm},
    \cite{della2025prompt}, \cite{gong2024well}, \cite{ICSEding2024vulnerability},
    \cite{ICSEyong2025gvi}, \cite{Tdaneshvar2025vulscriber}, \cite{Tchen2025security},
    \cite{JSScao2025enhancing}, \cite{FSELostintheEnd}, \cite{fanllm},
    \cite{ICSMEcamporese2025repairing} & 16 \\ \cline{2-4}
  & \textbf{GPT-3.5 Family} (Turbo, ChatGPT) &
    \cite{ICSEding2024vulnerability}, \cite{Tdaneshvar2025vulscriber},
    \cite{Tchen2025security}, \cite{ni2026learning}, \cite{JSSyang2025dlap},
    \cite{JSScao2025enhancing}, \cite{FSELostintheEnd}, \cite{fanllm},
    \cite{ICSMEcamporese2025repairing}, \cite{zhao2025towards},
    \cite{tony2025prompting}, \cite{tony2025discrete}, \cite{gong2024well}
    & 13 \\ \cline{2-4}
  & \textbf{Claude Family} (Opus, 3.5 Sonnet) &
    \cite{tianenhanced}, \cite{zhao2025towards} & 2 \\ \cline{2-4}
  & \textbf{Gemini Family} (1.5 Flash, Pro) &
    \cite{tony2025discrete}, \cite{tony2025retrieve} & 2 \\ \cline{2-4}
  & \textbf{Reasoning Models} (OpenAI o1/o3) &
    \cite{wen2025vul}, \cite{fanllm} & 2 \\ \hline

\multirow{4}{*}{\rotatebox[origin=c]{90}{\parbox{1.6cm}{Open Source}}}%
  & \textbf{Llama / CodeLlama} (7B, 13B, 70B) &
    \cite{TSE_CGP-Tuning}, \cite{mao2025towards}, \cite{Tchen2025security},
    \cite{FSELostintheEnd}, \cite{FSEran2025efficient}, \cite{tianenhanced},
    \cite{zhao2025towards}, \cite{tony2025discrete}, \cite{gong2024well}
    & 9 \\ \cline{2-4}
  & \textbf{DeepSeek-Coder} (V2, V2.5, R1-Distill) &
    \cite{mao2025towards}, \cite{Tchen2025security}, \cite{zhang2026vultrlm},
    \cite{fanllm}, \cite{tianenhanced}, \cite{ICSMEcamporese2025repairing},
    \cite{zhao2025towards}, \cite{tony2025discrete}
    & 8 \\ \cline{2-4}
  & \textbf{Qwen / CodeQwen} (1.5, 2.5, Coder) &
    \cite{Tdaneshvar2025vulscriber}, \cite{zhang2026vultrlm},
    \cite{FSEran2025efficient}, \cite{wen2025vul}, \cite{zhao2025towards}
    & 5 \\ \cline{2-4}
  & \textbf{StarCoder / 2} &
    \cite{ICSEding2024vulnerability}, \cite{li2025improving},
    \cite{wu2025mystique} & 3 \\ \hline

\multirow{4}{*}{\rotatebox[origin=c]{90}{\parbox{1.6cm}{Encoder-Only}}}%
  & \textbf{CodeBERT} &
    \cite{ICSEding2024vulnerability}, \cite{Tdaneshvar2025vulscriber},
    \cite{haque2026zero}, \cite{JSSkalouptsoglou2025transfer},
    \cite{FSEMoE}, \cite{ESEMkatz2025siexvults}, \cite{JSSren2025improving}
    & 7 \\ \cline{2-4}
  & \textbf{UniXcoder} &
    \cite{ICSEding2024vulnerability}, \cite{zhang2026vultrlm},
    \cite{haque2026zero}, \cite{FSEMoE} & 4 \\ \cline{2-4}
  & \textbf{GraphCodeBERT} &
    \cite{JSSren2025improving}, \cite{ESEMkatz2025siexvults} & 2 \\ \cline{2-4}
  & \textbf{SecureFalcon} &
    \cite{ferrag2025securefalcon} & 1 \\ \hline

\end{tabular}}
\end{table}
\renewcommand{\arraystretch}{1} 
\emph{Retrieval-augmented generation (RAG)}. 
We identified a small number of studies that employ RAG, typically relying on static retrieval sources such as CVE databases. However, the literature lacks RAG systems in which the retrieval knowledge base is dynamically updated based on detection or generation outcomes, particularly through human feedback.

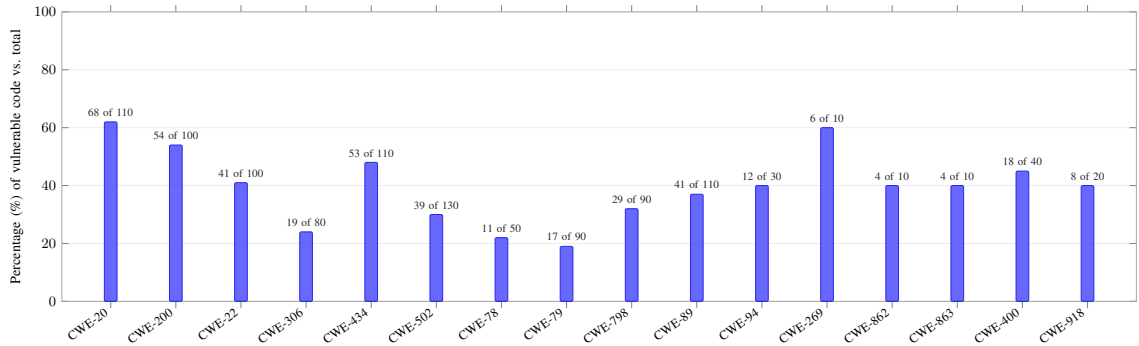
\begin{figure*}[t]
\centering
\scalebox{0.60}{%
\begin{tikzpicture}

\pgfplotstableread[row sep=\\]{
x       y      label      \\
CWE-20  62   68~of~110    \\
CWE-200 54   54~of~100    \\
CWE-22  41   41~of~100    \\
CWE-306 24   19~of~80     \\
CWE-434 48   53~of~110    \\
CWE-502 30   39~of~130    \\
CWE-78  22   11~of~50     \\
CWE-79  19   17~of~90     \\
CWE-798 32   29~of~90     \\
CWE-89  37   41~of~110    \\
CWE-94  40   12~of~30     \\
CWE-269 60   6~of~10      \\
CWE-862 40   4~of~10      \\
CWE-863 40   4~of~10      \\
CWE-400 45   18~of~40     \\
CWE-918 40   8~of~20      \\
}\cwetable

\begin{axis}[
    ybar,
    bar width=8pt,
    width=1.4\textwidth,
    height=8cm,
    ymin=0, ymax=100,
    enlarge x limits=0.05,
    ymajorgrids=true,
    grid style={gray!15},
    axis line style={gray!70},
    tick label style={font=\small},
    label style={font=\small},
    ylabel={Percentage (\%) of vulnerable code vs. total},
    symbolic x coords={
        CWE-20,CWE-200,CWE-22,CWE-306,CWE-434,CWE-502,CWE-78,
        CWE-79,CWE-798,CWE-89,CWE-94,CWE-269,CWE-862,CWE-863,CWE-400,CWE-918},
    xtick=data,
    x tick label style={
        rotate=35,
        anchor=east,
        font=\footnotesize
    },
]

\addplot[
    fill=blue!70,
    draw=blue!90!black,
    opacity=0.85,
    rounded corners=1pt,
    nodes near coords,
    nodes near coords style={font=\scriptsize, yshift=6pt},
    nodes near coords align={center},
    point meta=explicit symbolic
] table[x=x, y=y, meta=label]{\cwetable};


\end{axis}
\end{tikzpicture}}
\caption{Distribution of targeted CWEs in the ground-truth dataset}
\label{fig:CWE-scores}
\end{figure*}

\boxit{
Human feedback is typically applied only in the moment and does not influence future assessments.}
\phantom{To identify opportunities to improve the security of LLM-generated code, assess the reliability of results from static analysis tools, and determine }\\
\section{Manual vs. Automated Analysis Tools}
\label{sec:study2}

To identify opportunities to improve the security of LLM-generated code, assess the reliability of results from static analysis tools, and determine whether static analysis tools and LLMs are sufficient on their own, we designed a study using the dataset from Bruni et al.~\cite{Bruni2025}. Their work uses two popular peer-reviewed datasets \emph{LLMSecEval} and  \emph{SecurityEval} to generate code with LLMs and to evaluate multiple prompting strategies across different GPT models in order to determine which is most effective. They also apply two widely adopted static analyzers, Semgrep and CodeQL, to assess the security of the generated code. Because their setup is comprehensive and representative, we use their data as our starting point.

\subsection{Methodology}


Previous work has shown that the RCI (Recursive Critiques Improvement) prompting technique with GPT-4o yields the most secure code~\cite{Bruni2025}.
We therefore use this configuration as our baseline. We then filter the tasks to those mapped to the CWE Top 25, resulting in 1,080 code samples (108 unique tasks, each prompted 10 times). The benchmarks they used are CWE-based: each task is designed to evaluate the model on a specific CWE, so each generated code sample is associated with exactly one target CWE.

Two trained reviewers (one Ph.D. student and one M.Sc. student) conducted the manual checking under the oversight of a faculty expert. One reviewer examined all 1{,}080 samples; a second reviewer independently examined a stratified random subset (one sample per task, for a total of 108 code samples). 
The reviewers labeled code samples for the presence or absence of each task’s target CWE.
They also examined the GPT-generated critiques corresponding to each code sample to uncover any recurring patterns.
Disagreements were resolved through discussion; any remaining disagreements were resolved by the faculty expert.
In the end, we achieved high inter-rater agreement (Cohen’s $\kappa = 0.94$).
This process resulted in a ground truth dataset of 1{,}080 vulnerable and secure code samples. 
Finally, we compared the reports of static analysis tools versus this dataset.

\subsection{Result}

We found that 39\% of the generated code samples were vulnerable and 61\% were secure.
We evaluated each code sample only with respect to its designated target CWE and did not investigate the presence of other, non-target CWEs. In other words, samples that we labeled as secure with respect to their specific target CWE may still contain other, non-target CWEs.
Figure \ref{fig:CWE-scores} presents the distribution vulnerable and secure samples for each target CWE.





We noted two main forms of self-critiques in the results for each task. \emph{General critique} offers broad guidance without concrete references to the snippet or CWEs (e.g., “use secure functions”, “sanitize inputs”), whereas \emph{specific critique} provides targeted, actionable feedback tied to the snippet (e.g., naming APIs or lines, citing the relevant CWE, and prescribing exact changes). 
Our observations indicated that specific self-critiques yielded more secure code than general self-critique: 64\% secure versus 56\%, respectively.

\boxit{
We observed that GPT provided more specific, domain-aware self-critiques in secure code.
}

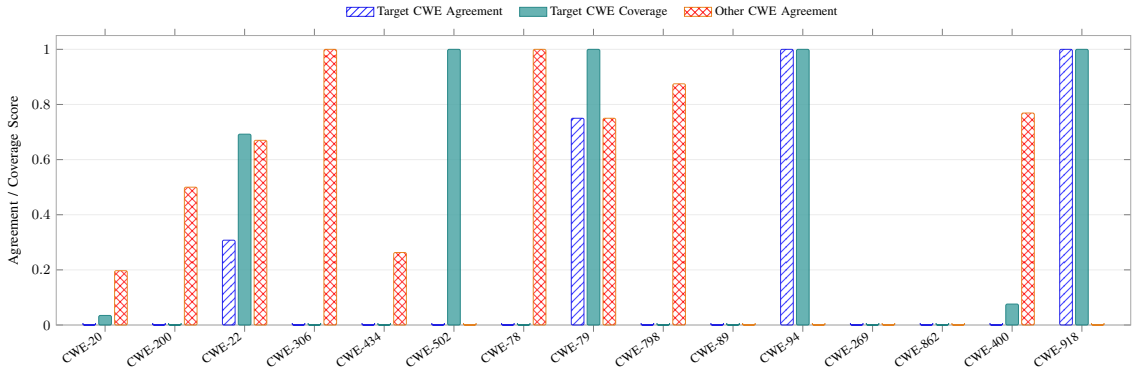
\begin{figure*}[t]
\centering
\scalebox{0.60}{\begin{tikzpicture}
\begin{axis}[
    ybar,
    bar width=8pt,
    width=1.4\textwidth,
    height=8cm,
    ymin=0, ymax=1.05,
    enlarge x limits=0.05,
    ymajorgrids=true,
    grid style={gray!15},
    axis line style={gray!70},
    tick label style={font=\small},
    label style={font=\small},
    ylabel={Agreement / Coverage Score},
    symbolic x coords={
        CWE-20,CWE-200,CWE-22,CWE-306,CWE-434,CWE-502,CWE-78,
        CWE-79,CWE-798,CWE-89,CWE-94,CWE-269,CWE-862,CWE-400,CWE-918},
    xtick=data,
    x tick label style={
        rotate=35,
        anchor=east,
        font=\footnotesize
    },
    legend style={
        font=\footnotesize,
        at={(0.5,1.04)},
        anchor=south,
        legend columns=3,
        draw=none,
        /tikz/every even column/.append style={column sep=0.3cm}
    },
    every node near coord/.append style={yshift=2pt}
]



\addlegendimage{area legend,pattern=north east lines,
pattern color=blue!70!blue,
  draw=blue!90!black,
opacity=0.85}
\addlegendentry{Target CWE Agreement}
\addlegendimage{area legend, fill=teal!70,
    draw=teal!90!black,
    opacity=0.85}
\addlegendentry{Target CWE Coverage}

\addlegendimage{area legend, pattern=crosshatch,
pattern color=red!70!red,
    draw=orange!90!black,
    opacity=0.85}
\addlegendentry{Other CWE Agreement}

\addplot[
pattern=north east lines,
pattern color=blue!70!blue,
  draw=blue!90!black,
opacity=0.85,
    rounded corners=1pt
] coordinates {
(CWE-20,0) (CWE-200,0) (CWE-22,0.3077) (CWE-306,0)
(CWE-434,0) (CWE-502,0) (CWE-78,0) (CWE-79,0.75)
(CWE-798,0) (CWE-89,0) (CWE-94,1) (CWE-269,0)
(CWE-862,0) (CWE-400,0) (CWE-918,1)
};


\addplot[
    fill=teal!70,
    draw=teal!90!black,
    opacity=0.85,
    rounded corners=1pt
] coordinates {
(CWE-20,0.0347) (CWE-200,0) (CWE-22,0.6923) (CWE-306,0)
(CWE-434,0) (CWE-502,1) (CWE-78,0) (CWE-79,1)
(CWE-798,0) (CWE-89,0) (CWE-94,1) (CWE-269,0)
(CWE-862,0) (CWE-400,0.076) (CWE-918,1)
};

\addplot[
pattern=crosshatch,
pattern color=red!70!red,
    draw=orange!90!black,
    opacity=0.85,
    rounded corners=1pt
] coordinates {
(CWE-20,0.197) (CWE-200,0.5) (CWE-22,0.67) (CWE-306,1)
(CWE-434,0.263) (CWE-502,0) (CWE-78,1) (CWE-79,0.75)
(CWE-798,0.875) (CWE-89,0) (CWE-94,0) (CWE-269,0)
(CWE-862,0) (CWE-400,0.769) (CWE-918,0)
};

\legend{
    Target CWE Agreement,
    Target CWE Coverage,
    Other CWE Agreement
}

\end{axis}
\end{tikzpicture}}

\caption{Tools agreement (intersection) and coverage (union) across CWEs based on ground truth}
\label{fig:Tools}
\end{figure*}




Semgrep and CodeQL flagged 60\% and 80\% of the samples as secure, respectively. Although Semgrep’s overall classification rate is close to the human assessment of 61\%, a per-sample comparison shows notable discrepancies, with only 65\% of Semgrep’s outputs and 61\% of CodeQL’s outputs agreeing with the ground-truth labels.
Table \ref{tab:static-tools-metrics} presents the performance of these tools based on our manual validation.

\begin{table}[ht]
\centering
\caption{Performance of static analysis tools}
\label{tab:static-tools-metrics}
\begin{tabular}{|l|c|c|c|}
\hline
\textbf{Tool} & \textbf{Recall} & \textbf{Precision} & \textbf{F-measure} \\ \hline
CodeQL           & 0.34 & 0.67 & 0.45 \\ \hline
Semgrep          & 0.54 & 0.52 & 0.53 \\ \hline
Both (agreement) & 0.30 & 0.76 & 0.43 \\ \hline
\end{tabular}
\end{table}

Figure \ref{fig:Tools} illustrates the tools' reports for each target CWE.
We observe that both tools fail to detect the target CWE in a substantial fraction of ground-truth cases, while frequently reporting findings in non-target CWE categories. The validity of these non-target findings remains unknown, as our ground-truth labels cover only the target CWE.




\boxit{
The tools' and human results were too different, highlighting the need for human feedback for reliable security assessment. 
}

\section{Proposed framework}
\label{sec:framework}

In RCI prompting technique, LLM attempts to fix issues raised during the critique step.
For example, the code in Listing~\ref{lst:redirect} represents an improved version of the model’s earlier output. While this version corrects several deficiencies, it still contains subtle security weaknesses. 
In particular, while the regular expression in line 3 restricts redirects to the \texttt{example.com} domain, the implementation remains susceptible to canonicalization or encoding-based bypasses, and the hostname extracted by \texttt{urlparse} is never validated directly (e.g., URLs such as \texttt{https://example.com@evil.com} or percent-encoded variants may still pass).

\begin{lstlisting}[language=Python,basicstyle=\ttfamily\scriptsize, breaklines=true, caption={Vulnerable LLM-generated code using RCI}, label={lst:redirect}]
import re
from urllib.parse import urlparse
URL_REGEX = re.compile(r"^https://(www\.)?example\.com(/[\w\-./?=#]*)?$")
DEFAULT_URL = "https://www.example.com/default"
def is_valid_url(url):
    parsed = urlparse(url)
    return (
        parsed.scheme == "https" and
        parsed.netloc and
        URL_REGEX.match(url)
    )
def get_redirect_url(url):
    return url if is_valid_url(url) else DEFAULT_URL
\end{lstlisting}

However, LLM frequently proposed this code piece as a secure sample. Static analysis tools also did not flag any risk.


When we provided contextual cues such as references to ``open redirect risks'' or ``percent-encoding bypasses'', the model then detected the remaining above mentioned security risks and fixed issues.


In summary, this study shows that human intervention is necessary to ensure code security. Humans can pinpoint areas where automated tools are wrong and provide extra context that improves LLM reasoning. We hypothesize that expert feedback is important not only for individual instances but may also generalize to similar cases.


\subsection{Framework}

\begin{figure*}[!h]
 	\centering
	\includegraphics[scale=0.27]{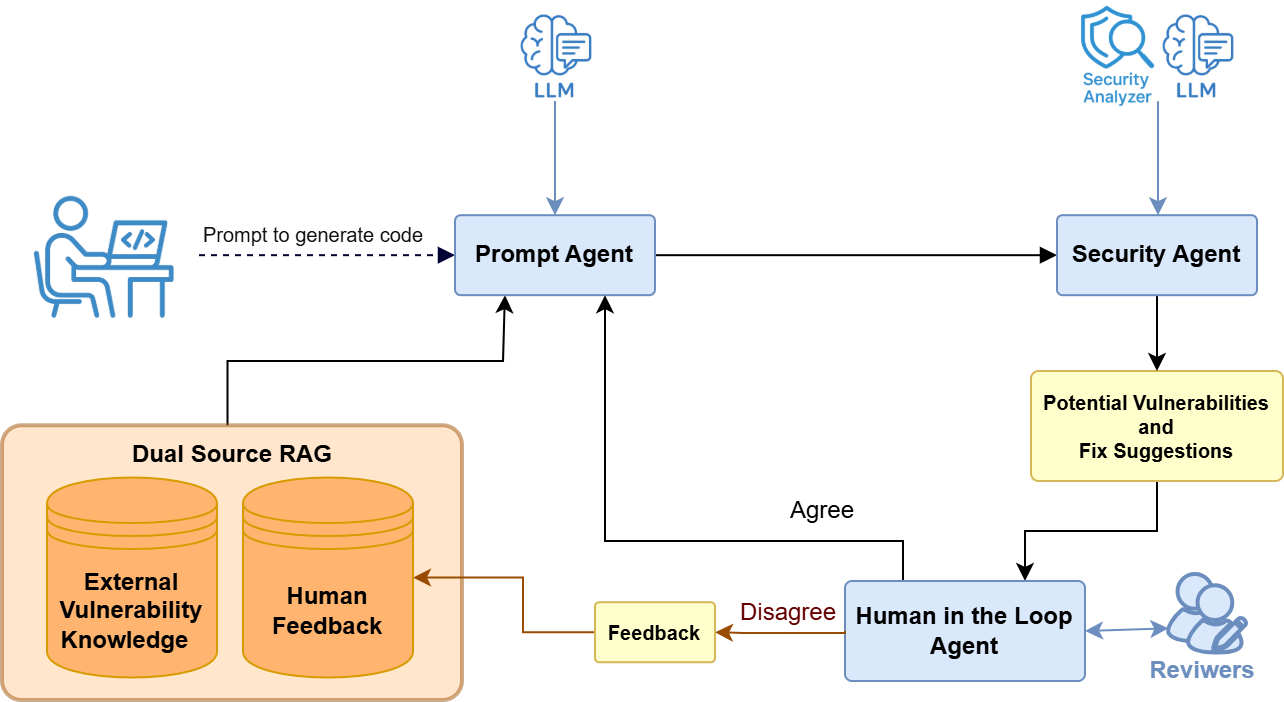}
	\caption{The proposed framework
    }
 	\label{fig:arch}   
\end{figure*}

LLMSecGuard~\cite{Kavian2024-lv} integrates traditional static analysis tools and LLMs to improve code security.
We propose a conceptual framework, illustrated in Figure~\ref{fig:arch}, that includes experts in this process and persistently stores
human feedback in a dynamic retrieval-augmented generation
pipeline, enabling LLMs to reuse past feedback for secure code
generation and vulnerability detection.


 \label{framework}

Starting from a developer prompt, the \emph{Prompt Agent} retrieves relevant security guidance from the RAG to improve the prompt and then synthesizes code with an LLM. The \emph{Security Agent} evaluates the resulting code using {security analysis tools}, and then passes the result to the \emph{Human (HIL) Agent}, where human experts review the Security Agent’s report. In cases of disagreement, the experts’ feedback is persistently stored in the RAG to cover similar future cases. 
The interplay between these agents continues until the decision of an expert.


\subsection{Prompt Agent}
It orchestrates code generation and refinement. Given a developer prompt, it forms an iteration-aware query that fuses the original request, and optional guidance from the HIL Agent. It queries the Dual-Source RAG and assembles a compact context under a token budget, then generates code using one or more LLMs. The agent follows an explicit update rule for iterative refinement:

\newpage

$$p_{i+1} = \mathrm{LLM}\!\bigl(\text{context}_i,\, p_i,\, h_i\bigr),$$
where $p_i$ is the current candidate and $h_i$ is optional HIL guidance. 

\subsection{Security Agent}

It evaluates each candidate program after each generation or refinement step. It runs static analyzers and leverages LLMs for vulnerability detection and explanation; together these components serve as the analysis engine. The agent generates a machine-readable report vector $f(p)$ (aggregating severity counts, residual-risk metrics, and suggested mitigations) and a human-readable report for experts. Both $f(p)$ and the report are routed to the HIL Agent. The Security Agent computes a score $s(p)$ to aid triage, but loop termination is decided exclusively by human experts via the HIL Agent.

\subsection{Human-in-the-Loop (HIL) Agent}
This agent is the final decision-maker. It provides expert reviews and curates safety knowledge. It enforces at least a \emph{dual review} process (Reviewer 1 and Reviewer 2) to guard against data poisoning and to reduce false positives and false negatives, and it returns concise, span-specific guidance $h_i$ for the next iteration. Whenever experts disagree with the Security Agent’s results, their feedback is stored in the RAG.

\textit{Trust Weight.}
When the HIL Agent promotes new guidance $d$ to the RAG, it assigns a trust weight $t(d)\in[0,1]$.  Unlike static scores, this weight is dynamic: it combines the immediate authority of human reviewers with the long-term empirical success of the guidance in real-world deployments.
$$t(d_{\text{new}}) \;=\; \min\!\left(1,\; t_{\text{base}} \;+\; \alpha\, N_{\text{rev}} \;+\; \beta\, s_d \right),$$
where $t_{\text{base}}\in[0,1)$ is the initial confidence, $N_{\text{rev}}\in\{1,2,\dots\}$ is the number of independent approvers (typically 2), and $\alpha,\beta\ge 0$ weight reviewer consensus and empirical performance. Items awaiting a second review are assigned $t(d)=0$ and excluded from default retrieval.

\textit{Dynamic Success Update (EMA).}
The term $s_d\in[0,1]$ is the historical success rate: how often this guidance led to HIL‑accepted code without further edits. Initialize $s_d^{(0)}=s_0$ (e.g., $s_0=0.5$). After the $n$‑th use,
$$s_d^{(n)} \;\leftarrow\; (1-\eta)\, s_d^{(n-1)} \;+\; \eta \cdot \mathbb{I}\{\text{HIL‑approved and criteria } \theta \text{ satisfied}\},$$
with learning rate $\eta\in(0,1]$ and indicator $\mathbb{I}\{\cdot\}\in\{0,1\}$. Here, $\theta$ encodes configurable automated acceptance checks (e.g., zero Critical/High findings, policy compliance, passing tests) used for accounting; HIL approval still gates loop termination.
This makes trust a living metric that adapts to real‑world performance while preserving human precedence.\\

\subsection{Dual-Source RAG}
A shared knowledge service consumed by the Prompt Agent. It exposes a single retrieval API over two distinct corpora: $HF$ (Human Feedback: human‑vetted via HIL dual review) and $EVK$ (External Vulnerability Knowledge: CVEs, vendor advisories, vulnerability patterns). Each document $d$ contains unstructured text $\text{text}(d)$, metadata $\text{meta}(d)$, a source label $\text{src}(d) \in \{\mathrm{HF}, \mathrm{EVK}\}$, and a trust weight $t(d) \in [0,1]$ assigned by the HIL workflow.
Retrieval relies on a composite scoring function. For a query $q$, the score $S(q,d)$ is defined as:
$$S(q,d) = \underbrace{\pi\bigl(\text{src}(d)\bigr)}_{\text{Source Priority}} \cdot \underbrace{t(d)}_{\text{Trust Weight}} \cdot \underbrace{R(q,d)}_{\text{Semantic Similarity}},$$
where $R(q,d) \ge 0$ is a relevance functional combining semantic and lexical signals. The source priority $\pi(\cdot)$ satisfies $\pi(\mathrm{HF}) > \pi(\mathrm{EVK})$ to strictly \emph{favor dynamic (human‑vetted) items}, since expert feedback should have higher priority than general guidance.
Selection follows a Dynamic‑Prioritized Backfill strategy. The system first selects up to $k$ dynamic items exceeding a quality threshold $\theta_{\mathrm{HF}}$, and fills remaining slots from the general corpus:

$$R_{\mathrm{HF}} = \operatorname{Top}_k\Bigl(\bigl\{\, d \in HF : S(q,d) \ge \theta_{\mathrm{HF}} \,\bigr\}\Bigr),$$
$$R_k = R_{\mathrm{HF}} \cup \operatorname{Top}_{k-|R_{\mathrm{HF}}|}\Bigl(\bigl\{\, d \in EVK : S(q,d) \ge \theta_{\mathrm{EVK}} \,\bigr\}\Bigr),$$
where $\operatorname{Top}_k(\cdot)$ returns the $k$ highest‑scoring items and $|R_{\mathrm{HF}}|$ is the cardinality of the dynamic set.

To resolve contradictions (e.g., general advice conflicting with expert feedback), the system applies a penalty:
 $$S(q, d_{\mathrm{EVK}}) \leftarrow \kappa \cdot S(q, d_{\mathrm{EVK}}), \quad \kappa \in (0,1),$$
The score of the general item is reduced by a factor $\kappa$, between 0 and 1 (e.g., 0.5). Subsequently, the conflicting pair is flagged for expert review.\\

To mitigate the feedback‑ingestion attack surface introduced by HF (e.g., data poisoning and prompt‑injection content) and to reduce false positives and false negatives, we enforce at least dual reviews. New or updated feedback items enter a staging pool with $t(d)=0$  and remain ineligible for serving until two distinct reviewers approve them: a primary reviewer and an independent approver.
\section{Threats to Validity}
\label{sec:threats}

\emph{Construct validity.}
Our literature review is restricted to 2025 publications in top-tier software engineering venues; earlier work and security-focused venues (e.g., USENIX Security, CCS, S\&P) may show different patterns.

\emph{Internal validity.}
Ground-truth labels rely on manual assessment and are thus subject to human bias despite high inter-rater agreement and expert oversight.

\emph{External validity.}
Our experiments focus on prompts from the LLMSecEval and SecurityEval datasets for Python, along with code generated using GPT‑4o with RCI prompting.
This mirrors prior work but limits generalization to other languages (especially low-level ones), domains, and large real-world systems. We also evaluate only one LLM (GPT-4o) and two static analyzers (CodeQL, Semgrep); using other LLMs or static analysis tools might yield different recall, precision, and CWE coverage.

\section{Conclusion}
\label{sec:Conclusion}

This work shows that benchmarks for secure code generation and vulnerability detection in the software engineering community are heavily concentrated on Python and C/C++. In addition, the widespread use of static security analysis tools without human validation to evaluate LLM outputs poses significant risks to the reliability of studies in this domain. 
We find that expert feedback is important not only for individual instances but may also generalize to similar cases. Accordingly, we propose a conceptual framework that persistently stores human feedback within a dynamic retrieval-augmented generation pipeline, enabling LLMs to reuse prior feedback for secure code generation and vulnerability detection.

\bibliographystyle{IEEEtran}
\bibliography{sample-base}
\end{document}